\RequirePackage{lineno}
\documentclass[aapm,mph,print,unsortedaddress,showkeys]{revtex4-2}
\usepackage[pdftex]{graphicx}
\usepackage{bm}
\usepackage[colorlinks=true,linkcolor=blue,citecolor=blue]{hyperref}
\usepackage[export]{adjustbox}
\usepackage{lineno,hyperref}
\usepackage{url}
\usepackage{amsmath}
\usepackage{array,booktabs}
\usepackage{color}
\modulolinenumbers[1]
\usepackage{amssymb}
\usepackage{float}
\usepackage{appendix}
\usepackage[utf8]{inputenc} % set input encoding (not needed with XeLaTeX)
\usepackage{caption}
\usepackage{wrapfig}
\usepackage{mathtools}
\usepackage{booktabs} % for much better looking tables
\usepackage{array} % for better arrays (eg matrices) in maths
\usepackage{paralist} % very flexible & customisable lists (eg. enumerate/itemize, etc.)
\usepackage{verbatim} % adds environment for commenting out blocks of text & for better verbatim
\usepackage{subcaption} % make it possible to include more than one captioned figure/table in a single float
\newcommand{\prob}{\mathrm{pr}}

\begin{document}

\def\papertitle{Invertibility of Multi-Energy X-ray Transform}
\title{\papertitle}

\author{Yijun Ding}
\email[Email: ]{dingy@email.arizona.edu}
\thanks{Address: 408 W. Simpson St., Tucson, AZ, 85701. Author to whom correspondence should be addressed.}
\affiliation{Wyant College of Optical Sciences, University of Arizona, Tucson, AZ, United States, 85719}

\author{Eric W.\ Clarkson}
\email[Email: ]{clarkson@radiology.arizona.edu}
\affiliation{Department of Medical Imaging and Wyant College of Optical Sciences, University of Arizona, Tucson, AZ, United States, 85719}

\author{Amit Ashok}
\email[Email: ]{ashoka@optics.arizona.edu}
\affiliation{Wyant College of Optical Sciences and Department of Electrical and Computer Engineering, Tucson, AZ, United States, 85719}

\date{\today}

Running title: Invertibility of ME X-ray transform

%%%%%%%%%%%%%%%%%% abstract %%%%%%%%%%%%%%%%
\begin{abstract}% <= 500 words

{\textbf{Purpose:}}  The goal is to provide a sufficient condition for the invertibility of a multi-energy (ME) X-ray transform. The energy-dependent X-ray attenuation profiles can be represented by a set of coefficients using the Alvarez-Macovski (AM) method. An ME X-ray transform is a mapping from $N$ AM coefficients to $N$ noise-free energy-weighted measurements, where $N\geq2$. 

{\textbf{Methods:}}  We apply a general invertibility theorem to prove the equivalence of global and local invertibility for an ME X-ray transform. We explore the global invertibility through testing whether the Jacobian of the mapping $J(\bm A)$ has zero values over the support of the mapping. The Jacobian of an arbitrary ME X-ray transform is an integration over all spectral measurements. A sufficient condition for $J(\bm A)\neq0$ for all $\bm A$ is that the integrand of $J(\bm A)$ is $\geq0$ (or $\leq0$) everywhere. Note that the trivial case of the integrand equals 0 everywhere is ignored. Using symmetry, we simplified the integrand of the Jacobian to three factors that are determined by the total attenuation, the basis functions, and the energy-weighting functions, respectively. The factor related to the total attenuation is always positive, hence the invertibility of the X-ray transform can be determined by testing the signs of the other two factors. Furthermore, we use the Cram\'er-Rao lower bound (CRLB) to characterize the noise-induced estimation uncertainty and provide a maximum-likelihood (ML) estimator.

{\textbf{Results:}} The factor related to the basis functions is always negative when the photoelectric/Compton/Rayleigh basis functions are used and K-edge materials are not considered. The sign of the energy-weighting factor depends on the system source spectra and the detector response functions. For four special types of X-ray detectors, the sign of this factor stays the same over the integration range. Therefore, when these four types of detectors are used for imaging non-K-edge materials, the ME X-ray transform is globally invertible. The same framework can be used to study an arbitrary ME X-ray imaging system, e,g, when K-edge materials are present. Furthermore, the ML estimator we presented is an unbiased, efficient estimator and can be used for a wide range of scenes.

{\textbf{Conclusions:}}  We have provided a framework to study the invertibility of an arbitrary ME X-ray transform and proved the global invertibility for four types of systems.

\end{abstract}

%%%%%%%%%%%%% keywords %%%%%%%%%%%%%%%%%%%%%%%%
\maketitle

%\linenumbers\modulolinenumbers[1]
%%%%%%%%%%%%%%  Introduction  %%%%%%%%%%%%%%%%%%%%%%%
\section{Introduction}
\indent Muti-energy (ME) X-ray imaging, also referred to as spectral or energy-selective X-ray imaging, has long been used to image the chemical composition of the object being scanned \cite{alvarez1976energy, lehmann1981generalized, barnes1985detector, cardinal1988theoretical, roessl2007k,fredenberg2018spectral}. In X-ray imaging, the chemical composition of a material is characterized by the energy dependence of the X-ray attenuation profile. As an X-ray attenuation profile can be represented as a linear combination of basis functions with known energy dependences, it can be summarized by a few energy-independent coefficients as in the Alvarez-Macovski method \cite{alvarez1976energy}. We refer to these coefficients as AM coefficients. Imaging AM coefficients requires multiple energy-weighted measurements, e.g. energy integration with varying source tube voltages or photo-counting with multiple energy bins. We refer to the mapping from the AM coefficients to the energy-weighted measurements an X-ray transform. The question whether an X-ray transform is invertible has only been explored recently for dual-energy (DE) \cite{levine2017nonuniqueness, alvarez2019invertibility} and ME measurements \cite{bal2020uniqueness}. The purpose of this work is to provide a sufficient condition for the invertibility of a general ME X-ray transform from a different perspective.

\indent With the recent developments in detectors, ME X-ray imaging is becoming more tangible. DE X-ray imaging recovers two AM coefficients \cite{alvarez1976energy} that represent contributions from photoelectric absorption and Compton scattering to the linear attenuation profile, respectively. The contribution from Rayleigh scattering has been considered negligible or assumed to be captured by the other two AM coefficients in DE X-ray imaging \cite{sukovle1999basis, lehmann1981energy, macovski1976energy}. However, it is difficult to predict the effect caused by ignoring the Rayleigh scattering term due to the nonlinear nature of the X-ray transform, especially for security- and industrial-screening applications where the materials of interest are not necessarily low-Z materials. With ME detectors, the AM coefficient corresponding to the Rayleigh scattering can be recovered. Furthermore, ME X-ray imaging systems can image materials containing K-edges in the spectral range used for imaging \cite{roessl2007k}. 

\indent With broad-spectrum X-ray sources, measurements of many X-ray systems are naturally energy-weighted \cite{tapiovaara1985snr}. ME measurements can be acquired with varying source settings \cite{flohr2006first, zhang2011objective} or with detectors with varying energy responses, such as sandwich detectors \cite{altman2009tu}, counting and integrating X-ray (CIX) detectors \cite{kraft2007counting} and multi-bin photon-counting detectors \cite{karg2005using}. More specifically, the recent advancement in photon counting detectors with pulse-height analysis, which output signals in multiple energy levels, provides a paradigm shift in X-ray detector technology and is enabling many new applications \cite{taguchi2013vision, fredette2019multi}. 

\indent The invertibility of a transform is a fundamental question in inverse problems. The invertibility problem considers noise-free measurements and determines whether a unique solution exists. A system of $N$ linear equations of $N$ unknowns has a unique solution (as long as the forward matrix is invertible); this is not necessarily true for nonlinear transforms. Levine et.\ al.\ \cite{levine2017nonuniqueness} demonstrated a case of DE X-ray imaging with non-unique solutions. Alvarez et.\ al.\ \cite{alvarez2019invertibility} has applied a two-dimensional global inverse theorem \cite{fulks1961advanced} to DE X-ray transforms. Bal et.\ al.\ \cite{bal2020uniqueness} provided an invertibility criteria for an ME X-ray transform by placing strong orientation constraints on the Jacobian matrices and demonstrated the equivalence of global and local invertibility for some examples through numerical experiments. We apply a global inverse function theorem for a N-dimensional map and prove that, for an ME X-ray transform, local invertibility is equivalent to global invertibility. Our global invertibility criteria is local invertibility, which is a weaker sufficient condition than the criteria provided by Bal et.\ al.\ \cite{bal2020uniqueness}. This is proved through topological properties of the definition region and inequalities of the X-ray transform. Furthermore, we provide a sufficient condition for the global invertibility by taking advantage of the symmetries in the expression of the Jacobian. With its simple expression, this condition can be applied to the design of ME X-ray imaging systems and detectors.

\indent In this paper, we provide a framework to study the invertibility of an arbitrary ME X-ray transform and prove the invertibility for four special cases of energy-weighted detectors. Furthermore, we consider Poisson noise in the measurement data and present the Cram\'er-Rao lower bound (CRLB) on the estimation of AM coefficients. Lastly, we provide a fast maximum-likelihood (ML) algorithm for coefficients estimation and demonstrate its application in an X-ray reconstruction problem.  

%%%%%%%%%%%%%%% forward model %%%%%%%%%%%%%%%%%%%%%%
\section{Forward problem: ME X-ray transform} \label{sec:meas_model}
\indent In the energy range 20---200 keV, which is commonly used for X-ray transmission imaging, the interaction between X-ray photons and the medium can be categorized into the following three processes: photoelectric absorption, Compton (incoherent) scattering and Rayleigh (coherent ) scattering. Correspondingly, the X-ray linear attenuation coefficient profiles can be represented accurately by a summation of $N$ terms as:
\begin{equation}
	\mu(E) = \sum_{i}^{N}a_{i}f_{i}(E) = {\bm a}\cdot{\bm f}(E),
\end{equation}
where each component of ${\bm f}$ is a function of energy $E$, the coefficients $\mathbf a$ are determined by the material composition, and the $N$ terms include photo electric, Compton scattering, Rayleigh scattering and K-edges. Here `photo electric' refers to the smooth energy dependence of the photo electric effect and `K-edges' refers to the discontinuities in the energy dependence of the photo electric effect just above the binding energy of the K-shell electrons. We use this set of $f_i(E)$ functions as basis functions and the coefficients $\bm a$ as the AM coefficients.

\begin{figure} [htbp!]
\centering
\includegraphics[width=\linewidth]{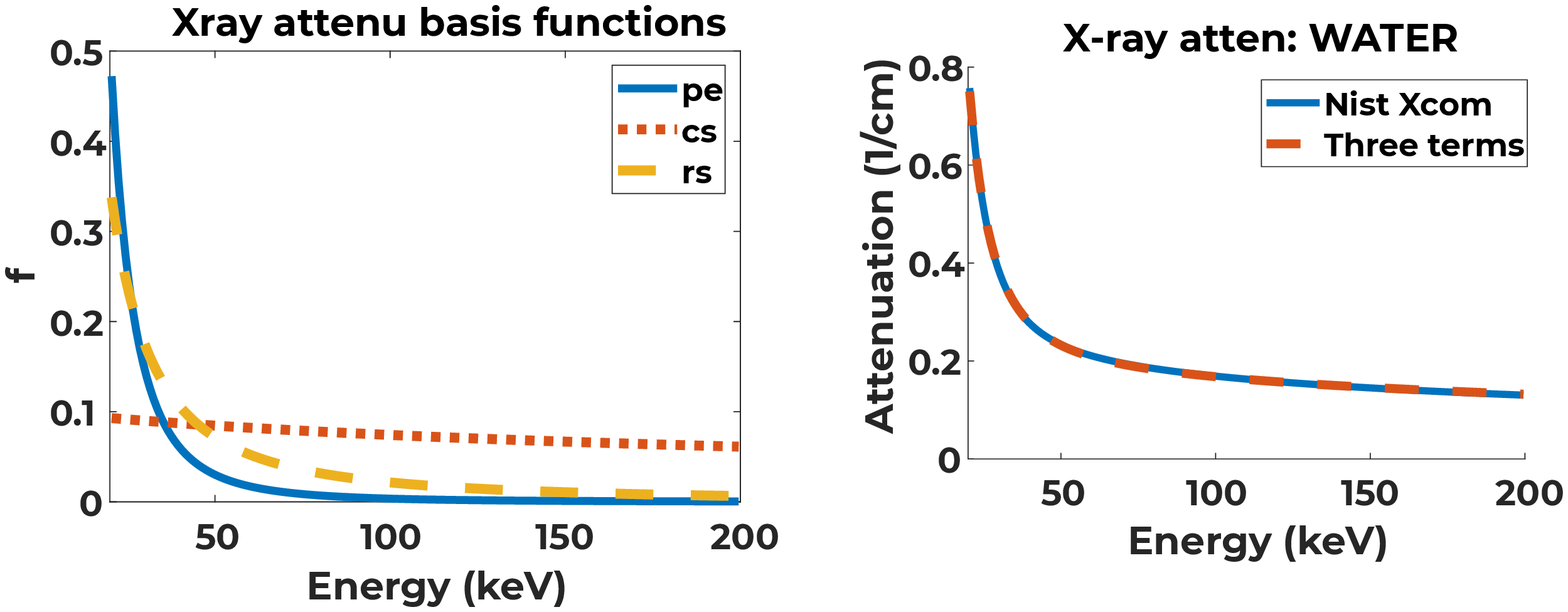}
\caption{Shape of $f_i(E)$ for $i=1,2,3$ (left) and fitted attenuation profile of water (right). }
\label{fig:basis_functions}
\end{figure}

\indent For materials that do not contain K-edges in the energy range of interest, the number of basis functions needed is $N=3$. Approximated expressions of photo electric and Rayleigh scattering term have been provided in Williamson et al.\ \cite{williamson2006two} by fitting to DLC-146 cross-section data \cite{trubey1989hugi} and the Klein-Nishina function \cite{alvarez1976energy} describes the Compton scattering term:
\begin{equation}
\begin{cases}
\begin{split}
	f_{1}(E) &= c_1E^{-3.088} ,\\
	f_{2}(E) &=c_2\left(\frac{1+\alpha}{\alpha^2}\left[\frac{2(1+\alpha)}{1+2\alpha}-\frac{1}{\alpha}\ln(1+2\alpha)\right]+\frac{1}{2\alpha}\ln(1+2\alpha)-\frac{1+3\alpha}{{(1+2\alpha)}^2}\right),\\
	f_{3}(E) &= c_3E^{-1.737},\\
	\end{split}
	\end{cases}
	\label{eq:f_terms}
\end{equation}
where $\alpha = E/(510.975~keV)$, the subscript 1,2,3 refers to the photoelectric effect, the Compton scattering and the Rayleigh scattering, respectively, and $c_i$ are normalization factors so that $\lVert f_i(E)\rVert_2=1$. The normalized basis functions are presented in Figure~\ref{fig:basis_functions} (left). The usefulness of these functions in representing attenuation coefficient profiles is well known. We generated attenuation profiles for 128 materials based on the NIST XCOM data \cite{berger1998photon}. As an example, the fitted attenuation profile and the XCOM data for water are presented in Figure~\ref{fig:basis_functions} (right). %Example of materials include Delrin, water, soap and cast magnesium. For each material, 1000 samples of slight different material compositions were generated. These attenuation profiles (K-edge materials excluded) were fitted by the three functions using a least-squares curve-fitting routine. Over the range of 20---200 keV, the largest difference between the fitted value and the XCOM value was less than $2.5\%$ with an average error of $0.66\%$. 

\indent In a tomographic imaging or measurement system, the total attenuation $\tau(E)$ is the line integral of the X-ray attenuation coefficient $\mu(E)$ along the ray path 
\begin{equation}
	\tau(E) = \int \mathrm{dl}~\mu(E)=\sum_{i=1}^N A_{i} f_{i}(E) = {\bm A}\cdot{\bm f}(E),
\end{equation}
where
\begin{equation}
	A_i = \int \mathrm{dl}~a_i(\bm R) 
\end{equation}
is a sinogram of the $i^{th}$ AM coefficient. For a parallel-beam system, $\bm A_i(\theta,\rho)$ is the Radon transform of $\bm a_i(\bm R)$, where $\theta$ is the rotation angle of the ray path and $\rho$ is the position along the detector plane. 

\indent The object $a_i(\bm R)$ can be reconstructed from the sinograms $\bm A_i(\theta, \rho)$, and the line integrals $\bm A_i(\theta,\rho)$ can be estimated from ME measurements of the corresponding ray path. Consider a ME X-ray imaging system producing $M$ energy-weighted measurements with a source photon budget $I_0$ (total number of photons emitted by the source across the energy range of interest). To describe the $m^{th}$ energy-weighted measurement, where $m=1,...,M$, denote $D_m(E)$ as the detector response and $S_m(E)$ as the normalized source spectrum of the $m^{th}$ measurement. For a given ray path, the mean signal of the $m^{th}$ measurement can be described by
\begin{equation}
\begin{split}
	I_m &= I_0\int_0^\infty \mathrm{dE}~D_m(E)\, S_m(E)\, \exp\left[-{\bm A}\cdot{\bm f}(E)\right],\\
	& =  I_0\int_0^\infty \mathrm{dE}~p_m(E)\, \exp\left[-{\bm A}\cdot{\bm f}(E)\right],\\
	\label{eq:mean_mapping}
	\end{split}
\end{equation}
where $p_m(E) = D_m(E)S_m(E)$ is the combined energy-weighting function. This equation can be used to describe many energy-weighted measurements, such as a photon-counting (PC) binning detector and an energy-integrating detector. In the most general case, the source spectra may vary across measurements and the combined weighting functions are arbitrary and can take on any real non-negative values at each energy. The basis functions $\bm f(E)$ can contain components describing K-edges as well. Therefore, Equation~(\ref{eq:mean_mapping}) describes a general ME X-ray transform. In the following sections, we study the invertibility of the mapping defined by this equation in the domain $A_i\geq0$ for $i=1,2,...,N$.

\begin{figure} [htbp!]
\centering
\includegraphics[width=\linewidth]{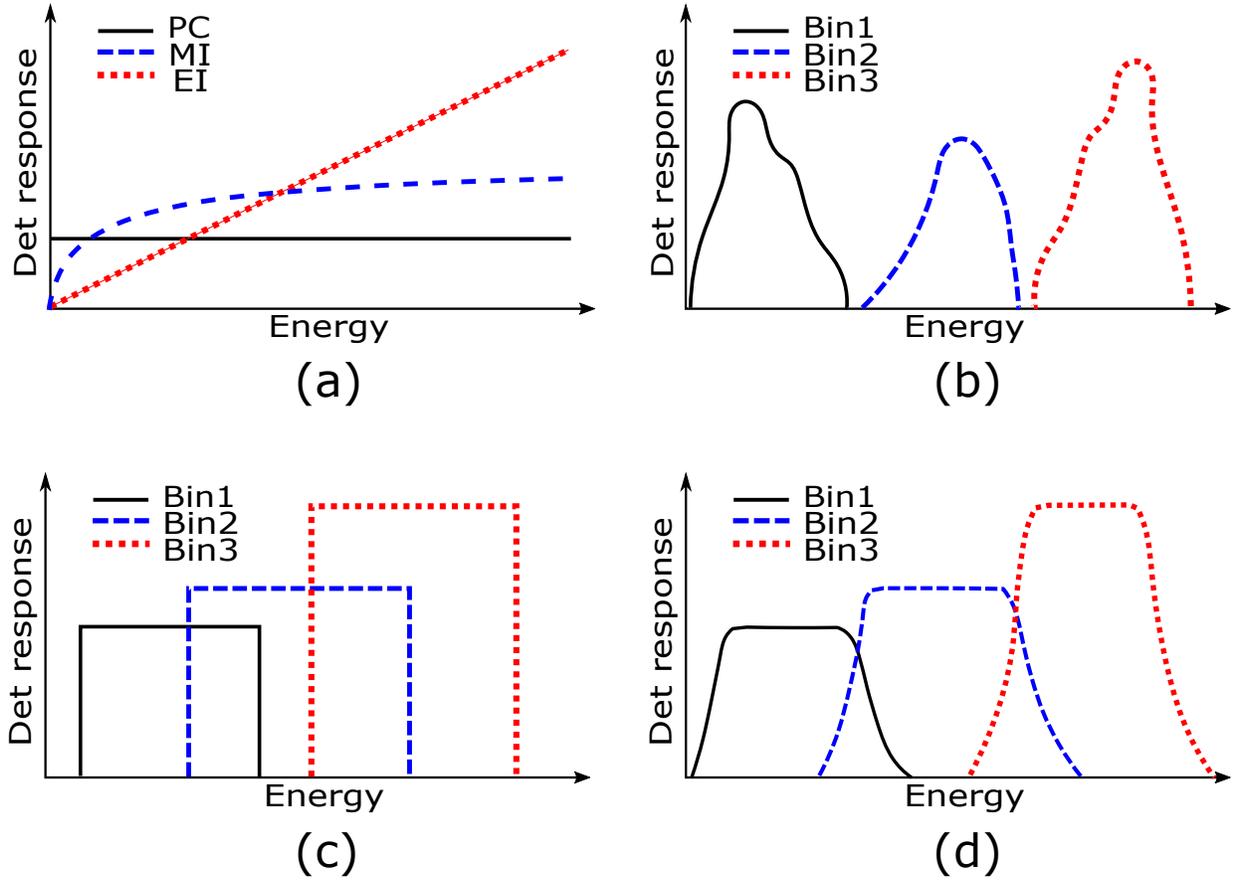}
\caption{Energy weighting functions for four special cases: (a) CIX-PC/EI/MI, (b)non-overlapping bins with arbitrary response, (c) three {\it rect} bins with ideal energy resolution, and (d) a slightly overlapping three-bin detector, where the overlap is introduced by the finite energy resolution of the detector. }
\label{fig:det_illu}
\end{figure}

\indent A special case of a ME detector is a CIX detector that counts the number of photons and integrates both the energy and the momentum of the photons (PC/EI/MI), providing measurements with detector response $D_1(E) = 1$, $D_2(E)\propto E$ and $D_3(E)\propto\sqrt{E}$ as shown in Figure~\ref{fig:det_illu}(a). As a CIX PC/EI detector has been developed \cite{kraft2007counting}, it is reasonable to assume that it is feasible to build a CIX PC/EI/MI detector. A second special case is a binning detector where the weighting functions are arbitrary and non-overlapping as shown in Figure~\ref{fig:det_illu}(b). Another special case is an ideal PC detector as illustrated in Figure~\ref{fig:det_illu}(c), where the detector response of each bin can be considered as $\it rect$ functions and there may be overlaps between different bins. Binning detectors in real life tend to have non-overlapping bins. Here, for the comprehensiveness, we include detectors with overlapping bins. Another special case considers a slightly overlapping three-bin detector, where the overlap is introduced by the finite energy resolution of the detector. The detector response functions of such a detector are plotted in Figure~\ref{fig:det_illu}(d).

%%%%%%%%%%%%%%%%%% invertibility %%%%%%%%%%%%%%%%%%%%

\section{Invertibility}
\indent We explore the invertibility of the mapping from the AM coefficients ${\bm A}$ to the noise-free ME measurement data ${\bm I}$. Suppose we have $M=N$ measurements. The coefficients ${\bm A}$ and the mean photon count ${\bm I}$ are both subsets of N-dimensional Euclidean space ${\bm R}^N$. We define the ME X-ray transform from ${\bm A}$ to ${\bm I}$ as $\mathcal{X}: M_1 \rightarrow M_2$, where the domain of the mapping is ${\bm A}$ in $M_1$ and the range of the mapping is ${\bm I}$ in $M_2$. 

\indent {\bf The Hadamard's global inverse function theorem}\cite{krantz2012implicit}: {\it Let $M_1$ and $M_2$ be smooth, connected N-dimensional manifolds and let $\mathcal{X}: M_1 \rightarrow M_2$ be a $C^1$ function. If (1) $\mathcal{X}$ is proper, (2) the Jacobian of $\mathcal{X}$ vanishes nowhere and (3) $M_2$ is simply connected, then $\mathcal{X}$ is a homeomorphism.} A homeomorphism is one-to-one and onto, which implies global invertibility, while non-vanishing Jacobian implies local invertibility. 

\indent In the following sections, we will use the Hadamard's global inverse function theorem to prove the equivalence of global and local invertibility for an ME X-ray transform. We will first construct a simply-connected range $M_2$, then prove the mapping $\mathcal{X}: M_1 \rightarrow M_2$ is proper through inequality relations. Lastly, we will provide a simplified expression for the Jacobian determinant and a sufficient condition for the Jacobian to vanish nowhere.

\subsection{Simply connected}
\indent We briefly summarize the property of the mapping $\mathcal{X}$. The first-order derivative of the X-ray transform can be expressed as follows:
\begin{equation}
\frac{\partial I_m}{\partial A_i} = - I_0\int_0^\infty \mathrm{dE}~p_m(E)\, f_i(E) \exp\left[-{\bm A}\cdot{\bm f}(E)\right].
\end{equation}
The first-order derivative exists and is continuous, therefore the mapping $\mathcal{X}$ is a $C^1$ mapping. %In fact, $\mathcal{X}$ is a $C^\infty$ mapping, as the mapping is differentiable for all degrees of differentiation. 
The values $\mathcal{X}(0)$, which represents the mean signals of an air scan, are finite and equal to the maximum (mean) count values. We further define a normalization factor $d_m=\int_{0}^{\infty}\mathrm{dE}~{p_{m}\left(E\right)}$. With this definition, the maximum mean count measured by the $m^{th}$ detector is $I_0d_m$. As the magnitude of $\bm A$ approaches infinity, the counts approach zeros. We will use these properties and the assumption that the Jacobian is non-vanishing in $\bm R^N$ to construct a simply connected range $M_2$ with the corresponding domain $M_1$ connected. Furthermore, we will justify that the interior of $M_1$ and $M_2$ are both smooth N-dimensional manifolds.

\indent The ME X-ray transform, as defined in Equation~(\ref{eq:mean_mapping}), has physical meaning when $\bm A$ is in the positive subspace of ${\bm R}^N$, denoted as ${\bm P}^N$, where $A_i\geq0$ for all $i$. However, the transform is mathematically valid over the domain ${\bm R}^N$. In order to construct a simply connected $M_2$, we will expand the domain of the mapping from ${\bm P}^N$.

\indent Let $U_0 = {\bm P}^N$ and $V_0$ be the image of $U_0$ under the mapping $\mathcal{X}$. In $V_0$, the mean photon count $\bm I$ is bounded by $0<I_m\leq I_0d_m$. Furthermore, as $U_0$ is path connected and $\mathcal{X}$ is a continuous mapping, $V_0$ is path connected  \cite{munkres2000topology}(page 150). From every point $\bm I_i$ in $V_0$, draw a straight line to the maximum-count point $I_0\bm d$ and define this line with end points as $V_i$. Every point in $V_i$ is bounded by $I_{im}\leq I_m\leq I_0d_m$. Define $M_2$ as the union of all $V_i$. As $V_0$ and $V_i$ are all path connected, $M_2$ is path connected. The space $M_2$ is simply connected if every closed curve in $M_2$ can be contracted to a point \cite{krantz2012implicit}. Define a closed curve $\phi(s):[0,1]\rightarrow M_2$. We can contract $\phi(s)$ to the maximum-count point $I_0\bm d$ through the following continuous function $H$: $[0,1]\times[0,1]\rightarrow M_2$, 
\begin{equation}
H(s, t) = t\phi(s) + (1-t)I_0\bm d. 
\label{eq:homotopy}
\end{equation}
As $M_2$ is path connected, the closed curve $\phi(s)$ can be contracted to any points in $M_2$ \cite{munkres2000topology} (page 332). Therefore, $M_2$ is simply connected.

\begin{figure} [htbp!]
\centering
\includegraphics[width=1\linewidth]{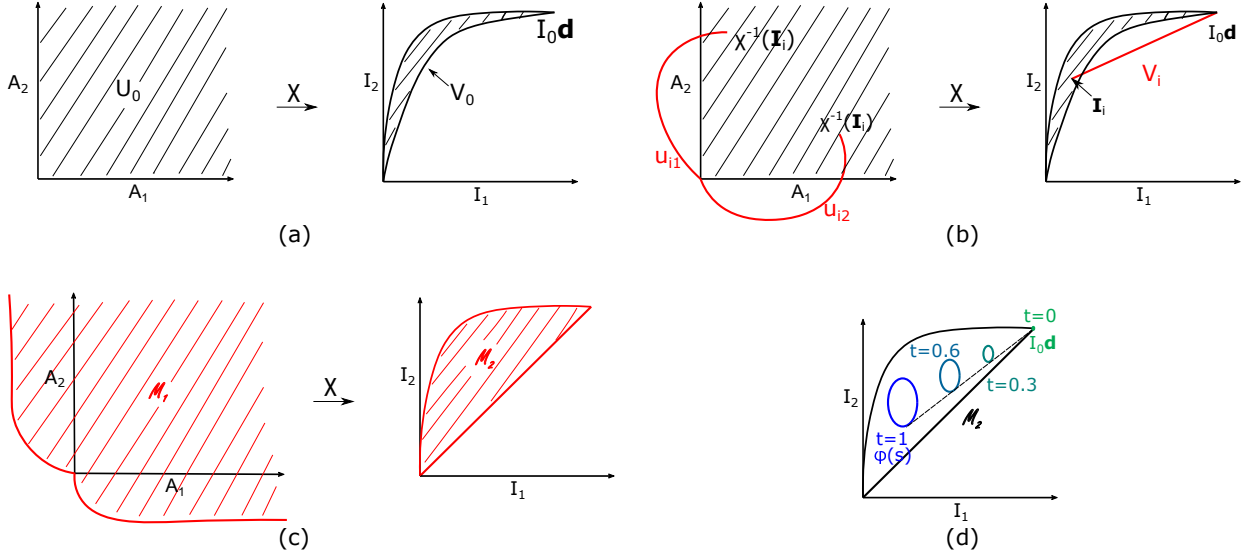}
\caption{Constructing a simply connected range $M_2$ by expanding the domain of the map $\mathcal{X}$: (a) The initial domain $U_0$ and range $V_0$. (b) $V_i$, defined by points $\bm I_i$ and $\bm I_0\bm d$, and its corresponding preimages $u_{ij}$. $U_i$ is defined as the union of all $u_{ij}$. (c) The expanded domain $M_1$ and range $M_2$. (d) Every closed curve $\phi(s)$ (blue loop) can be shrunk down to the point $I_0\bm{d}$ through the function $H(s,t)$. Color varying from blue to green represents $t$ from $1$ to 0. The black-dashed line is the trajectory of a point $s$ in the loop $\phi$ as $t$ decreases from 1 to 0. }
\label{fig:simply_connected_illu}
\end{figure}

\indent We assume that the Jacobian vanishes nowhere in ${\bm R}^N$. In other words, the mapping $\mathcal{X}$ is a local homeomorphism, which means that every point of $\bm A\in {\bm R}^N$ has a neighborhood that is homeomorphic to an open subset in the range. For every straight line $V_i(\bm I)$, the corresponding preimage $u_i(\bm A)$ in the domain ${\bm R}^N$ can be constructed by successive local inverses $\mathcal{X}^{-1}(\bm I)$. %As $u_i(\bm A)$ is the preimage of $V_i(\bm I)$ and a small neighborhood around it, $u_i(\bm A)$ is an open set. 
The maximum-count point $I_0\bm d$ corresponds to only one point in $U_0$, and this point is the origin of the coefficient space. Therefore, the origin is one end point of all preimages. The point $\bm I_i$ may have multiple local inverses, which we can index with subscript $j$. The $j^{th}$ local inverse introduces an inverse curve $u_{ij}$, where the $j^{th}$ local inverse is the second end point of the corresponding preimage $u_{ij}$. Each $u_{ij}$ is connected and connected to $U_0$. We define the union of all $u_{ij}(\bm A)$ as $U_i$. Furthermore, we define $M_1$ as the union of all $U_i$. $M_1$ is connected and a superset of ${\bm P}^N$. %Note that $U_i$ and $M_1$ are all open sets. 
The expansion of the range and domain for $N=2$ is illustrated in Figure~\ref{fig:simply_connected_illu}.

\indent The interior points of $M_1$ and $M_2$ (excluding the boundaries) are both smooth N-dimensional manifolds, because they are open subsets of ${\bm R}^N$ \cite{lee2012introduction} (page 19). We will limit our proof to the interior of $M_1$ and $M_2$ and discuss the boundary points in Section~\ref{sec:disc}.

\subsection{Proper}
\begin{figure} [htbp!]
\centering
\includegraphics[width=1\linewidth]{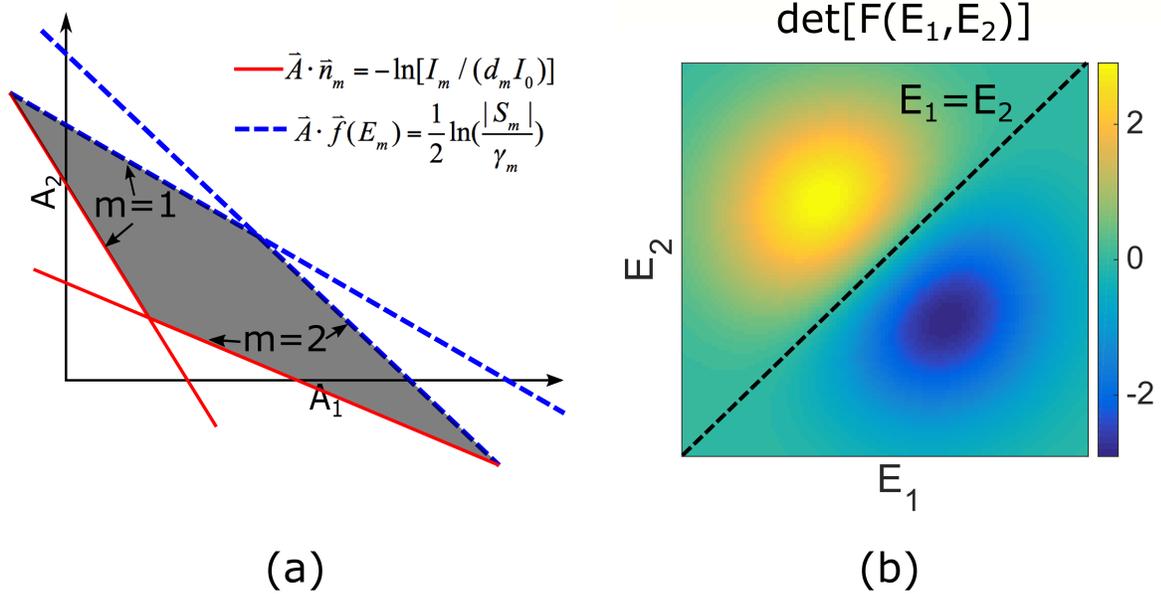}
\caption{(a) For given noise-free measurement data $\bm I$, the vector $\bm A$ is bounded in the area indicated in grey. Each energy-weighted measurement generates a pair of red and blue hyperplanes, which bound the vector $\bm A$. (b) Illustration of the sign-switched mirror symmetry in function $\det[F(E_1,E_2)$] along line $E_1=E_2$.}
\label{fig:eric_analysis}
\end{figure}

\indent We derive the bounds on the coefficients ${\bm A}$ for given measurement data ${\bm I}$. Using Jensen's inequality, we have
\begin{equation}
\ln \frac{I_{m}}{d_mI_0}>\int_{0}^{\infty}\mathrm{dE}~\frac{p_{m}\left(E\right)}{d_m}\left[-{\bm A}\cdot{\bm f}\left(E\right)\right],
\end{equation}
where $d_m$ has been defined previously and $d_mI_0$ is the maximum count in the $m^{th}$ measurement. These inequalities can be written as
\begin{equation}
{\bm A}\cdot{\bm n}_{m}>\ln[(d_mI_0)/I_{m}]
\label{eq:inequality1}
\end{equation}
where 
\begin{equation}
{\bm n}_{m}=\int_{0}^{\infty}\mathrm{dE}~p_{m}\left(E\right){\bm f}\left(E\right)/d_m
\end{equation}
The vector ${\bm n}_{m}$ has all non-negative components. Furthermore, the mean photon count $I_m$ in $M_2$, is always less than or equal to the maximum count, $d_mI_0$. Therefore, the right hand side of Equation~(\ref{eq:inequality1}) is always larger than or equal to 0. Each of the inequalities in Equation~(\ref{eq:inequality1}) forces the vector ${\bm A}$ to be on the side that is opposite to the origin of the hyperplane defined by
to ${\bm A}\cdot{\bm n}_{m}=-\ln[I_{m}/(d_mI_0)]$, as shown in Figure~\ref{fig:eric_analysis}(a) for the case of $N=2$. 

\indent We define the support of the weighting functions $p_{m}(E)$ as $\Omega_m$. Using the Schwarz inequality we have
\begin{equation}
I_{m}^{2}\leq I_0^2\left[\int_{\Omega_m}\mathrm{dE}~p_{m}^{2}\left(E\right)\right]\left[\int_{\Omega_m}\mathrm{dE}~\exp\left[-2{\bm A}\cdot{\bm f}\left(E\right)\right]\right],
\end{equation}
with equality if and only if $p_{m}(E)\propto\exp\left[-{\bm A}\cdot{\bm f}\left(E\right)\right]$. In many occasions, the equality condition is not attainable. For example, when the three basis functions given in Equation~(\ref{eq:f_terms}) are used, $\exp\left[-{\bm A}\cdot{\bm f}\left(E\right)\right]$ is not proportional to the $p_{m}(E)$ of the detectors illustrated in FIG. 2(b)-(d). If we define
\begin{equation}
\gamma_{m}=\frac{I_{m}^2}{I_0^2}\left[\int_{\Omega_m}~\mathrm{dE}~p_{m}^{2}\left(E\right)\right]^{-1},
\end{equation}
then we have
\begin{equation}
\gamma_{m}\leq\int_{\Omega_m}\mathrm{dE}~\exp\left[-2{\bm A}\cdot{\bm f}\left(E\right)\right].
\end{equation}
Assume that the length $\left|\Omega_m\right|$ of each support set is finite. Replacing the integrand with its maximum possible value
gives
\begin{equation}
\gamma_{m}\leq\exp\left\{ -2\min_{E\in \Omega_m}\left[{\bm A}\cdot{\bm f}\left(E\right)\right]\right\} \left|\Omega_m\right|
\end{equation}
Therefore we have another set of inequalities
\begin{equation}
\min_{E\in \Omega_m}\left[{\bm A}\cdot{\bm f}\left(E\right)\right]\leq\frac{1}{2}\ln\left(\frac{\left|\Omega_m\right|}{\gamma_{m}}\right)
\end{equation}
Now we may choose an energy $E_{m}$ such that
\begin{equation}
{\bm A}\cdot{\bm f}\left(E_{m}\right)\leq\ln\left(\sqrt{\frac{\left|\Omega_m\right|}{\gamma_{m}}}\right)
\label{eq:schwarz_bound}
\end{equation}
The right-hand side satisfies $\ln\left(\sqrt{\left|\Omega_m\right|/\gamma_{m}}\right)\geq\ln(d_mI_0/I_m)\geq0$ through the Schwarz inequality as follows:
\begin{equation}
d_m^2=\left[\int_{\Omega_{m}}\mathrm{dE}~p_{m}\left(E\right)\right]^{2}\leq |\Omega_m|\left[\int_{\Omega_{m}}\mathrm{dE}~p_{m}^{2}\left(E\right)\right].
\end{equation}
Each of the inequalities in Equation~(\ref{eq:schwarz_bound}) forces the vector ${\bm A}$ to be on the same side of the corresponding hyperplane as the origin. 

\indent Therefore, for given mean photon count ${\bm I}$, where $d_mI_0\geq I_m>0$, the inequalities defined by Equation~(\ref{eq:inequality1}) and (\ref{eq:schwarz_bound}) force ${\bm A}$ to be in a bounded set defined by the first set of hyperplanes and the second set of hyperplanes. A typical picture of this scenario for $N=2$ is shown in Figure~\ref{fig:eric_analysis}(a). Note that for a physical measurement, the corresponding coefficients $\bm A$ are further bounded by the coordinate planes. Here we focus on demonstrating that $\bm A$ is bounded for a given $\bm I$ even without the positivity constraints on $\bm A$.

\indent Now we show that the mapping $\mathcal{X}: M_1 \rightarrow M_2$ is a proper mapping. If we have a compact set $C$ in the data space $M_2$, where all of the data vectors are located, then there are maximum and minimum values for each $I_{m}$ over all ${\bm I}$ in $C$.
The maximum value for $I_{m}$ determines the hyperplane ${\bm A}\cdot{\bm n}_{m}=\ln[(d_mI_0)/I_{m}]$ that is closet to the origin. The minimum value for $I_{m}$ determines the hyperplane 
${\bm A}\cdot{\bm f}\left(E_{m}\right)=\ln\left(\sqrt{\left|\Omega_m\right|/\gamma_{m}}\right)$
that is furthest away from the origin. Therefore, the set of ${\bm A}$ that are mapped into $C$ are contained in a region bounded by these two sets of hyperplanes. This bounded region together with its boundary form  a closed and bounded set in $\mathbb{R}^{N}$, hence a compact set. As the map ${\mathcal{X}}\left({\bm A}\right)$ is continuous, the set of ${\bm A}$ that are mapped into the closed set $C$ is closed. The set of ${\bm A}$ that are mapped into $C$ is a closed subset of a compact set. This set is therefore also compact. As a result, the mapping ${\mathcal{X}}({\bm A})$ is proper.

\subsection{Jacobian}
\indent The Jacobian of the mapping is $J({\bm A}) = |\det(\nabla_{{\bm A}}{\bm I})|$, where $|\cdot|$ represents the absolute value and $\det(\cdot)$ is the determinant of a matrix. The matrix inside the determinant is
\begin{equation}
	\nabla_{{\bm A}}{\bm I} = \begin{bmatrix}
\frac{\partial I_1}{\partial A_1}  & \frac{\partial I_1}{\partial A_2}  & ... & \frac{\partial I_1}{\partial A_N} \\
\frac{\partial I_2}{\partial A_1}  & \frac{\partial I_2}{\partial A_2}  & ... & \frac{\partial I_2}{\partial A_N}\\
... & ... & ... & ...\\
\frac{\partial I_M}{\partial A_1}  & \frac{\partial I_M}{\partial A_2}  & ... & \frac{\partial I_M}{\partial A_N}
\end{bmatrix}.
\end{equation}
As $M=N$, $\nabla_{{\bm A}}{\bm I}$ is a square matrix. Defining the set $\Omega=\Omega_1\times \Omega_2...\times \Omega_m$, the determinant of a square matrix can be expressed in Leibniz formula as
\begin{equation}
\begin{split}
	\det(\nabla_{\bm A}{\bm I})&= \sum_{\bm\sigma}\text{sgn}(\bm\sigma)\prod_{m=1}^M\frac{\partial I_m}{\partial A_{\sigma_m}}  , \\
	&=(-I_0)^M\sum_{\bm\sigma}\text{sgn}(\bm\sigma)\prod_{m=1}^M\int_{\Omega_m}\mathrm{dE_m}~p_m(E_m)\, f_{\sigma_m}(E_m) \exp\left[-{\bm A}\cdot{\bm f}(E_m)\right]\\
	&= (-I_0)^M\int_{\Omega}\mathrm{d^M{\bf E}}~\left\{\prod_{m=1}^M p_m(E_m)e^{-{\bm A}\cdot{\bm f}(E_m)}\right\} \sum_{\bm\sigma}\text{sgn}(\bm\sigma)\prod_{m=1}^M f_{\sigma_m}(E_m), \\
	\end{split}
	\label{eq:jacobian_derivation}
\end{equation}
where the sum is computed over all permutations $\bm\sigma$ of the set $\{1,2,...,M\}$, and the sign of the permutation $\bm\sigma$, $\text{sgn}(\bm\sigma)$, is +1 or -1 for even or odd permutations, respectively.
Invoking the Leibniz formula again, we can simplify the Jacobian to:
\begin{equation}
\begin{split}
	\det(\nabla_{\bm A}{\bm I})&= (-I_0)^M\int_{\Omega}\mathrm{d^M{\bf E}}~\left\{\prod_{m=1}^M p_m(E_m)e^{-{\bm A}\cdot{\bm f}(E_m)}\right\} \det[F({\bm E})], \\
	&= (-I_0)^M\int_{\Omega}\mathrm{d^M{\bf E}}~\left\{\prod_{m=1}^M f_m(E_m)e^{-{\bm A}\cdot{\bm f}(E_m)}\right\} \det[P({\bm E})], \\
	\end{split}
	\label{eq:jacobian}
\end{equation}
where the second line follows a similar derivation as Equation~(\ref{eq:jacobian_derivation}) by swapping the subscripts of $p(E)$ and $f(E)$. The matrix $F$ as a function of ${\bm E} = (E_1, E_2, ..., E_M)$ is
\begin{equation}
	F({\bm E})  = \begin{bmatrix}
f_1(E_1) & f_2(E_1) & ... & f_N(E_1)\\
f_1(E_2) & f_2(E_2) & ... & f_N(E_2)\\
... & ... & ... & ...\\
f_1(E_M)&f_2(E_M) & ... & f_N(E_M)
\end{bmatrix},
\end{equation}
and the matrix $P$ as a function of ${\bm E}$ is
\begin{equation}
	P({\bm E})  = \begin{bmatrix}
p_1(E_1) & p_2(E_1) & ... & p_N(E_1)\\
p_1(E_2) & p_2(E_2) & ... & p_N(E_2)\\
... & ... & ... & ...\\
p_1(E_M)&p_2(E_M) & ... & p_N(E_M)
\end{bmatrix}.
	\label{eq:PE_matrix}
\end{equation}

\indent The integrand in Equation~(\ref{eq:jacobian}) has several interesting symmetry properties. The factor $\prod_{m=1}^M e^{-{\bm A}\cdot{\bm f}(E_m)}=e^{-{\bm A}\cdot{\sum_{m=1}^M \bm f}(E_m)}$ has mirror symmetry about all hyperplanes $E_i=E_j$ for $i,j\in\{1,2,..,M\}$. The other factor, $\det[F(\bm E)]$, has sign-switching mirror symmetry about the same hyperplanes, which can be described mathematically as:
\begin{equation}
\det[F({E_1, E_2, ..., E_M})] = \text{sgn}(\bm\sigma)\det[F(E_{\sigma_1}, E_{\sigma_2},..., E_{\sigma_M})].
\end{equation}
A sign-switching mirror symmetry means that, when we switch the positions of two coordinates (odd permutation), the sign of the function changes but the absolute value of the function is preserved. For example, with two coordinates $E_1$ and $E_2$, $\det[F({E_1, E_2})]=f_1(E_1)f_2(E_2)-f_2(E_1)f_1(E_2) =  -\det[F({E_2, E_1})]$. To illustrate the sign-switching mirror symmetry, we plotted $\det[F(E_1,E_2)]$ for the case when $f_1(E)$ and $f_2(E)$ are both Gaussian functions in Figure~\ref{fig:eric_analysis}(b). 

\indent Now we can divide the space occupied by $\Omega$ into $M!$ subspaces with hyperplanes $E_i=E_j$ for $i,j\in\{1,2,..,M\}$. One of the subspace has property $E_1<E_2<...<E_M$ and we define this subspace as $\Omega_{1,2...M}$. For every point $(E_1, E_2, ..., E_M)$ in the subspace $\Omega_{1,2...M}$, there is a corresponding point $(E_{\sigma_1}, E_{\sigma_2}, ..., E_{\sigma_M})$ in each of the remaining subspaces. Applying the sign-switching mirror symmetry of the determinant, we can further simplify the Jacobian to:
\begin{equation}
\begin{split}
	\det(\nabla_{\bm A}{\bm I})&= (-I_0)^M\int_{\Omega_{12..M}}\mathrm{d^M{\bf E}}~\left\{\prod_{m=1}^M e^{-{\bm A}\cdot{\bm f}(E_m)}\right\} \det[F({\bf E})] \sum_{\bm\sigma}\text{sgn}(\bm\sigma)\prod_{m=1}^Mp_m(E_{\sigma_m}), \\
	&=(-I_0)^M\int_{\Omega_{12..M}}\mathrm{d^M{\bf E}}~\left\{\prod_{m=1}^M e^{-{\bm A}\cdot{\bm f}(E_m)}\right\} \det[F({\bm E})] \det[P(\bm E)].\\
	\end{split}
	\label{eq:jacobian_subspace}
\end{equation}
The symbol $M$ emphasizes that the integration is over all spectral measurements. The three factors of the integrand, $\prod_{m=1}^M e^{-{\bm A}\cdot{\bm f}(E_m)}$, $\det[F({\bm E})]$ and $\det[P(\bm E)]$ depend on the total attenuation, the basis functions and the energy-weighting functions, respectively. 

\indent  When the Jacobian is non-vanishing everywhere in ${\bm R}^N$, we can construct the domain $M_1$ and the range $M_2$ and prove that the mapping $\mathcal{X}: M_1\rightarrow M_2$ is globally invertible. As a result, the ME X-ray transform defined on the domain $P^N$, which is a subset of $M_1$, is globally invertible. Therefore, we have proved the equivalence of global invertibility with local invertibility for an ME X-ray transform. This equivalence holds when K-edge basis functions are considered.

\subsection{A sufficient condition for invertibility}
A sufficient but not necessary condition for $J(\bm A)\neq0$ is that the integrand of $J(\bm A)$, which is given in Equation~(\ref{eq:jacobian_subspace}), has the same sign over the subspace $S_{12...N}$ and has non-zero values. The first factor in the integrand, which depends solely on the total attenuation, is always positive. If we ignore the trivial case that $\det[F({\bm E})]\det[P(\bm E)]=0$ everywhere, {\bf a sufficient condition for the invertibility of the ME X-ray transform is $\det[F({\bm E})]\det[P(\bm E)]\leq0$ (or $\geq0$) for all $\bm E$ in $S_{12...N}$ }. As the sign of the integrand does not depend on $\bm A$, if the sufficient condition is satisfied, the Jacobian is non-vanishing for all $
\bm A$ in ${\bm R}^N$.

\indent When the photoelectric/Compton/Rayleigh basis functions are used, the basis-function determinant $\det[F({\bm E})]$ is always negative in the subspace $\Omega_{123}$. This set of three basis functions is sufficient to describe an object when the materials of interest have no K-edges in the energy range used for imaging, e.g. soft tissue and bone. The proof of $\det[F({\bm E})]<0$ for all $\bm E$ that satisfy $E_1<E_2<E_3$ is provided in Appendix~A. When K-edge materials are considered, the values of $\det[F({\bm E})]$ can be calculated numerically and the positive regions of $\det[F({\bm E})]$ can be avoided by adjusting the detector sensitivity or source spectrum in $p_m(E)$.

\indent Now we apply the sufficient condition for invertibility to the DE scenario that has non-unique solutions discussed by Levine \cite{levine2017nonuniqueness}. For DE X-ray imaging, a set of two basis functions, photoelectric and Compton, can be used. The basis-function determinant is always positive in the subspace $S_{12}$. Levine assumed the same detector response for the two measurements, hence the sign of $\det[P({\bm E})]$ is the same with the sign of $\det[S({\bm E})]$, which is the source-spectra determinant. Similar to the definition in Equation~(\ref{eq:PE_matrix}), the matrix $S({\bm E})$ can be written as
\begin{equation}
	S(E_1, E_2)  = \begin{bmatrix}
S_1(E_1) & S_2(E_1) \\
S_1(E_2) & S_2(E_2) \\
\end{bmatrix},
	\label{eq:SE_matrix}
\end{equation}
where $(E_1, E_2)$ can be any combinations of two energies. With these, $\det[S({\bm E})] = S_1(E_1)S_2(E_2)-S_1(E_2)S_2(E_1)$. Levine assumed that both source spectra $S_1(E)$ and $S_2(E)$ are not zero only at three energy points (30, 60, 100)~keV. Hence, $\det[S({\bm E})]$ is not zero only when $(E_1, E_2)$ are combinations of the set $\{30, 60, 100\}$~keV. Within the subspace $S_{12}$, where $E_1$ is always less than $E_2$, $\det[S({\bm E})]$ is non-vanishing only at three points $(E_1, E_2) = (30, 60)$, $ (30, 100)$, and $ (60, 100)$~keV. Given the two source spectra as $S_1(E) =(1, 1, 1)$ and $S_2(E) = (0.93, 1.71, 0.30)$ at $E=(30, 60, 100)$~keV, respectively. The values of $\det[S({\bm E})]$ at $(E_1, E_2) = (30, 60)$, $ (30, 100)$, and $ (60, 100)$~keV are 0.78, -0.63 and -1.41, respectively. Therefore, $\det[S({\bm E})]$ does not have constant sign over the subspace $S_{12...N}$, hence the invertibility of the DE X-ray transform for the proposed scenario is not guaranteed. Note that this analysis only shows that the existence of non-unique solutions is possible, it does not prove their existence. 

\indent We then consider the sign of $\det[P({\bm E})]$ for the four types of detectors illustrated in Figure~\ref{fig:det_illu}. Assuming the source spectrum $S(E)$ is same for different $m$, the weighting-function determinant $\det[P({\bm E})]$ has the same sign with the detector-response determinant $\det[D({\bm E})]$, where the $(i,j)$ element of matrix $D({\bm E})$ is the detector response of the $i^{th}$ measurement at $E_j$ denoted as $D_i(E_j)$. The sign of $\det[D({\bm E})]$ for the four types of detectors are studied in Appendix~A and the main results are presented as follows. 

\renewcommand{\theenumi}{\alph{enumi}}
\renewcommand{\labelenumi}{(\theenumi)}
\begin{enumerate}
     \item An CIX-PC/EI/MI detector: $\det[D({\bm E})]<0$ for all $\bm E\in \Omega_{123}$.
     \item A three bin detector, where the three bins are not overlapping and the energy-response functions are arbitrary: $\det[D({\bm E})]\geq0$ for all $\bm E\in \Omega_{123}$. 
     \item A three bin PC detector with {\it rect}-response functions and possible overlaps between bins: if Bin 1 and Bin 3 has no overlap, the lower edges of the three bins satisfy $l_1<l_2<l_3$ and the upper edges of the three bins satisfy $u_1<u_2,<u_3$, the determinant $\det[D({\bm E})]\geq0$ for all $\bm E\in \Omega_{123}$.
     \item A non-overlapping three bin PC detector with finite energy resolution: if the energy resolution of the detector can be modeled by a narrow truncated function (for mathematical description see Appendix~A) and there is no overlap between Bin 1 and Bin 3,  $\det[D({\bm E})]\geq0$ in $\Omega_{123}$. 
\end{enumerate}

\indent In conclusion, the mapping $\mathcal{X}: {\bm A}\rightarrow {\bm I}$ is globally invertible for these four types of detectors when measuring attenuation profiles without K-edges. For arbitrary detectors or systems with varying source spectra, the values of $\det[P({\bm E})]$ can be calculated numerically. One can alway maintain $\det[F({\bm E})]\det[P({\bm E})]\leq 0$ for any $E$ in $S_{12...N}$ by adjusting the energy-response function or the bin boundaries of the detectors.

%%%%%%%%%%%%%%%%%% uncertainties %%%%%%%%%%%%%%%%%%%%

\section{Estimation uncertainties for Poisson data} \label{sec:crb}

\indent From a practical point of view, it is also crucial to consider the uncertainty in the estimation under the presence of noise. In this section, we consider PC detectors with non-overlapping bins. If only inherent quantum noise is considered, the data of the $m^{th}$ measurement at a given ray path, $g_m$, is a Poisson random variable with mean equals to $I_m$,
\begin{equation}
	g_m(\bm A) = {\mathcal Poiss}(I_m(\bm A)),
\end{equation}
where $I_m(\bm A)$ is the mean photon count of the $m^{th}$ measurement given in Equation~(\ref{eq:mean_mapping}). Combining all $M$ measurements, we get the measurement data ${\bm g}$. The probability density function of the data ${\bm g}$ given the AM coefficient $\bm A$ along the ray path is 
\begin{equation}
\prob{({\bm g}|{\bm A})} = \prod_{m=1}^M\frac{I_m({\bm A})^{g_m}e^{-I_m({\bm A})}}{g_m!}
\label{eq:data_pdf_poisson}
\end{equation}

\indent The log-likelihood function of the AM coefficients $\bm A$ is
\begin{equation}
	L({\bf A}|\bm g) =\ln{\prob({{\bm g}|{\bm A}})}= \sum_{m=1}^M g_m\ln{I_m({\bf A}}) - I_m(\bm A) - \ln{g_m!}.
	\label{eq:likelihood_poisson}
\end{equation}
The first derivative of the log-likelihood function is
\begin{equation}
	\frac{\partial L}{\partial A_i}(\bm A) = \sum_{m=1}^M \frac{g_m-I_m(\bm A)}{I_m(\bm A)}\frac{\partial I_m}{\partial A_i}(\bm A) = \sum_{m=1}^M \frac{g_m-I_m(\bm A)}{I_m(\bm A)}[\nabla_{\bm A}{\bm I}]_{mi}.
\end{equation}
The Hessian, or second derivative, of the log-likelihood function is given by
\begin{equation}
	[\nabla^2_{\bm A}L]_{ij} = \frac{\partial^2 L}{\partial A_i\partial A_j} = \sum_{m=1}^M \frac{g_m-I_m}{I_m}\frac{\partial^2 I_m}{\partial A_i\partial A_j} - \frac{g_m}{I_m^2}\frac{\partial I_m}{\partial A_i}\frac{\partial I_m}{\partial A_j}.
\end{equation}

\indent The components of the Fisher information matrix are
\begin{equation}
\begin{split}
	\mathrm{FIM}_{ij}({\bm A}) &= -\left<\frac{\partial^2L}{\partial A_i \partial A_j} \right>_{{\bm g}|{\bm A}} \\
	& = -\int \mathrm{d^Mg}~\prob({\bm g}|{\bm A})\sum_{m=1}^M\left[\frac{g_m-I_m}{I_m}\frac{\partial^2I_m}{\partial A_i\partial A_j} - \frac{\partial I_m}{\partial A_i}\frac{\partial I_m}{\partial A_j}\frac{g_m}{I_m^2}\right] \\
	& = \sum_{m=1}^M \frac{1}{I_m}\frac{\partial I_m}{\partial A_i}\frac{\partial I_m}{\partial A_j}.
	\end{split}
\end{equation}
Therefore, the Fisher information matrix is 
\begin{equation}
	\mathrm{FIM}({\bm A}) = (\nabla_{\bm A}{\bm I})^T \Lambda^{-1}(\nabla_{\bm A}{\bm I}),
\end{equation}
where $\Lambda$ is a diagonal matrix with the $m^{th}$ diagonal element equals to $I_m$. 

\indent The {\it Cr\'amer-Rao bounds} \cite{cramer1946mathematical, rao1945information} characterize the limit on the estimation uncertainties induced by noise. It states that for an unbiased estimate of the $i^{th}$ parameter, its variance must be at least as large as the $i^{th}$ diagonal element of the inverse of the Fisher information matrix. Mathematically, the Cr\'amer-Rao lower bounds are
\begin{equation}
	\text{Var}(\hat A_i-A_i)\geq[\mathrm{FIM}^{-1}]_{ii} = [(\nabla_{\bm A}{\bm I})^{-1} \Lambda(\nabla_{\bm A}{\bm I})^{-1,T}]_{ii},
\end{equation}
where the symbol $\hat A$ indicates an estimate of $A$. Note that the uncertainty in the estimation is inversely related with the source photon budget $I_0$, which agrees with our intuition. Also note that the uncertainty of an unbiased estimation is inversely related with the Jacobian $J(\bm A)=|\det(\nabla_{\bm A}{\bm I})|$. If the Jacobian $J(\bm A)$ is close to zero, the estimation uncertainty is close to infinity and the coefficients can not be estimated accurately in practice. 

%%%%%%%%%%%%%%%%%% demonstration %%%%%%%%%%%%%%%%%%%%

\section{Estimation algorithm and illustrative results}

\indent In this section we develop a maximum-likelihood (ML) algorithm for Poisson data. The goal of the algorithm is to estimate AM-coefficients $\bm A$ from noisy data $\bm g$. The assumption for the algorithm is that both Equations~(\ref{eq:mean_mapping}) and ~(\ref{eq:data_pdf_poisson}) are valid. First, consider $L$ as a function of the mean signal $\bm I$. The first derivation of this function is
\begin{equation}
	\frac{\partial L}{\partial I_i} = \frac{g_i}{I_i}-1.
\end{equation}
Hence, the point $\bm I=\bm g$ is a critical point. The Hessian of the function $L(\bm I|\bm g)$ is
\begin{equation}
	[\nabla^2_{\bm I}L]_{ij} = \frac{\partial^2 L}{\partial I_i\partial I_j} = -\frac{g_i}{I_i^2}\sigma_{ij}
\end{equation}
where $\sigma_{ij}=1$ when $i=j$, and $\sigma_{ij}=0$ otherwise. This Hessian is a diagonal matrix with all negative elements when $g_m>0$. Therefore, the function $L(\bm I|\bm g)$ is a concave function and the critical point at $\bm I=\bm g$ is the global maximum for $L$. When the mapping $\mathcal{X}: {\bm A}\rightarrow {\bm I}$ is invertible and $\bm g$ is within the range of the mapping, the maximum value of $L$ corresponds to a point $\bm A_{g}$ that satisfies $\bm I(\bm A_g) = \bm g$. 

\indent We then consider $L$ as a function of the AM coefficients $\bm A$. When the matrix $\nabla_{\bm A}{\bm I}$ is invertible, which is true when the X-ray transform is invertible, $\bm A_g$ is a critical point of the likelihood function $L(\bm A|\bm g)$, as $L(\bm A_g|\bm g)$ is the maximum likelihood value. Furthermore, when $\bm A$ is located within the region defined by $I_m(\bm A)\geq g_m$, the likelihood function $L(\bm A|\bm g)$ is a concave function of $\bm A$. An ML algorithm can be developed based on Newton's method \cite{boyd2004convex, nocedal2006numerical} with iterations described as
\begin{equation}
	\bm A_{k+1} = \bm A_{k} + t_k*\Delta\bm A_k,\quad\quad {\text and} \quad\quad \Delta\bm A_k=-[\nabla^2_{\bm A}L(\bm A_k)]^{-1}\nabla_{\bm A}L(\bm A_k),
\end{equation}
where $\bm A_{k}$ is the attenuation coefficients at iteration $k$ and $t_k$ is the step size chosen with an Armijo-type (or back-tracking) line search to enforce sufficient increase in $L$ and the negativeness of the Hessian $\nabla^2_{\bm A}L$. Note that the enforcement of negative-definiteness of the Hessian is important, as the algorithm may not converge otherwise. The likelihood function is convex in the region $0\leq\bm A\leq \bm A_g$, hence the algorithm works the best when the initialization point $\bm A_{0}$ is less than $\bm A_g$. Furthermore, during the iterations, the Hessian may become rank-deficient due to numerical accuracy. In this case, a gradient-descent step can be used instead of the Newton step. For example, in our implementation, we used an initialization of $\bm A_{0}=(0,0,0)$ and back-tracking parameters $\alpha=0.1$ and $\beta=0.1$ ($\alpha$ and $\beta$ are defined as in \cite{boyd2004convex}), and the convergence of the algorithm required less than 15 iteration steps for every case we have tested in the following sections.  

\subsection{AM coefficients estimation uncertainties}

\indent To demonstrate the applications of this maximum-likelihood algorithm, we considered an ideal three-bin PC detector with non-overlapping {\it rect} response functions, as shown in Fig~2(c) but with equal heights and no overlapping. The source is operated at 160 kVp and generates a broad X-ray spectrum \cite{ding2019x}. The material attenuation profiles are extracted from the NIST XCOM data. X-ray attenuation was simulated according to Beer's law. X-ray scattering and detector imperfections are not considered. The data were simulated at source photon budget $I_0=10^7$ with Poisson noise. 

\begin{figure} [htbp!]
\centering
\includegraphics[width=1\linewidth]{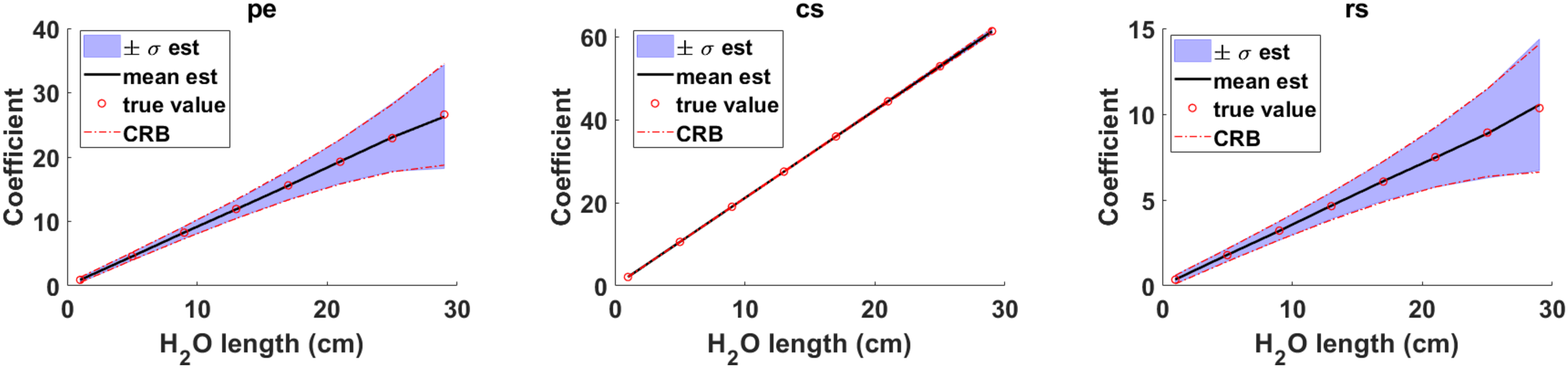}
\caption{Uncertainties in the three estimated photoelectric/Compton-scattering/Rayleigh-scattering coefficients for different lengths of water. }
\label{fig:water_est}
\end{figure}

\indent We simulated a single X-ray path with different lengths of water as the attenuating media. The energy-bin boundaries of the detector were set at [30, 75, 100, 160]~keV. The length of water ranged from 1~cm to 28~cm. For each length, 1000 sets of noisy data were generated, and the AM coefficients were estimated for each set of data. We calculated the mean and the variance of the estimated coefficients and compared the estimation uncertainty to the Cr\'amer-Rao lower bound. The results are presented in Figure~\ref{fig:water_est}.

\begin{figure} [htbp!]
\centering
\includegraphics[width=1\linewidth]{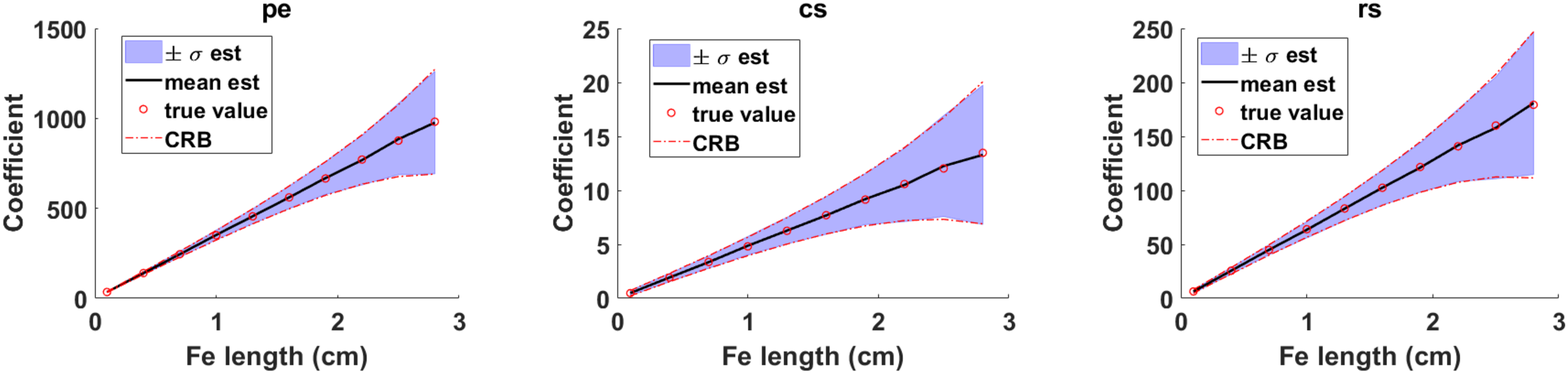}
\caption{Uncertainties in the three estimated photoelectric/Compton-scattering/Rayleigh-scattering coefficients for different lengths of iron. }
\label{fig:Fe_est}
\end{figure}

\indent To check if the algorithm works for materials that are very different from water, we changed the material in the X-ray path to iron and the detector bin boundaries to [30, 110, 140, 160]~keV. The bin boundaries were changed so that there are sufficient number of photons collected in all three bins. The length of iron ranged from 0.1~cm to 3~cm. The mean and variance of the 1000 repeated estimations at each length are presented in Figure~\ref{fig:Fe_est}.

\indent In both scenarios, the mean of the estimates (black line) matches well with the true coefficient (red circle). The slight deviation between the true coefficient and the mean estimation at high attenuation region can be attributed to sampling error in Monte-Carlo simulation. The standard deviation of the estimates (purple area) is almost perfectly aligned with the Cr\'amer-Rao lower bound. These results demonstrate that our estimation algorithm is unbiased and efficient. Furthermore, the AM coefficient corresponding to the Rayleigh scattering is estimable and the uncertainty in $\hat A_{rs}$  is comparable to the uncertainty in $\hat A_{pe}$, which corresponds to the photoelectric effect. However, the $\hat A_{pe}$ and $\hat A_{rs}$ are anti-correlated, which is demonstrated by a negative value of the element $[\mathrm{FIM}^{-1}]_{13}$. This correlation is probably due to the fact that the basis functions $f_{pe}(E)$ and $f_{rs}(E)$ have similar shapes. This correlation may partially explain why $\hat A_{pe}$ and $\hat A_{rs}$ have more variance than $\hat A_{cs}$, as shown in Figure~\ref{fig:water_est}. As a result of the correlation, the estimated total attenuation $\hat\tau(E)=\hat{A}_{pe}f_{pe}(E)+ \hat{A}_{cs}f_{cs}(E)+\hat{A}_{rs}f_{rs}(E)$, which is the ultimate physical quantity we are interest in, is not very noisy, as will be shown in the reconstruction results of the phantom. 

\subsection{Phantom reconstruction}

\indent We further applied the ML estimation algorithm for an image reconstruction. The reconstruction problem in X-ray computed tomography (CT) is to estimate the distribution $\bm a(\bm R)$ from the estimated line integrals $\hat{\bm A}$ for each ray path. The AM coefficients $\bm a$ at each location $\bm R$ correspond to an attenuation profile $\mu(E)$. For a two-dimensional scene, the object $\mu(E,\bm R)$ and the reconstruction $\hat{\mu}(E,\bm R)$ are both three-dimensional data cubes. To present the reconstruction result, we plot $\mu(E,\bm R)$ and $\hat{\mu}(E,\bm R)$ at an arbitrary energy.

\indent We simulate a two-dimensional fan-beam CT system (62$^\circ$ fan-beam angle) with 360 views and 245 detectors. The field of view is 256x256 pixels with a pixel pitch of 1---1.5~mm. The same source, detector, and material database described in the previous section are used. X-ray attenuation is simulated according to Beer's law, while scattering and detector imperfections are not considered. The detector energy bin boundaries are [30, 75, 100, 160]~keV. The source photon budget for each beam path is $10^7$. The AM coefficients $\bm A$ are estimated for each beam path and the object represented by $\bm a(\bm R)$ is reconstructed from $\hat{\bm A}$ using a filtered-back projection (FBP) algorithm. 

\begin{figure}[htbp!]
\centering
\includegraphics[width=1\linewidth]{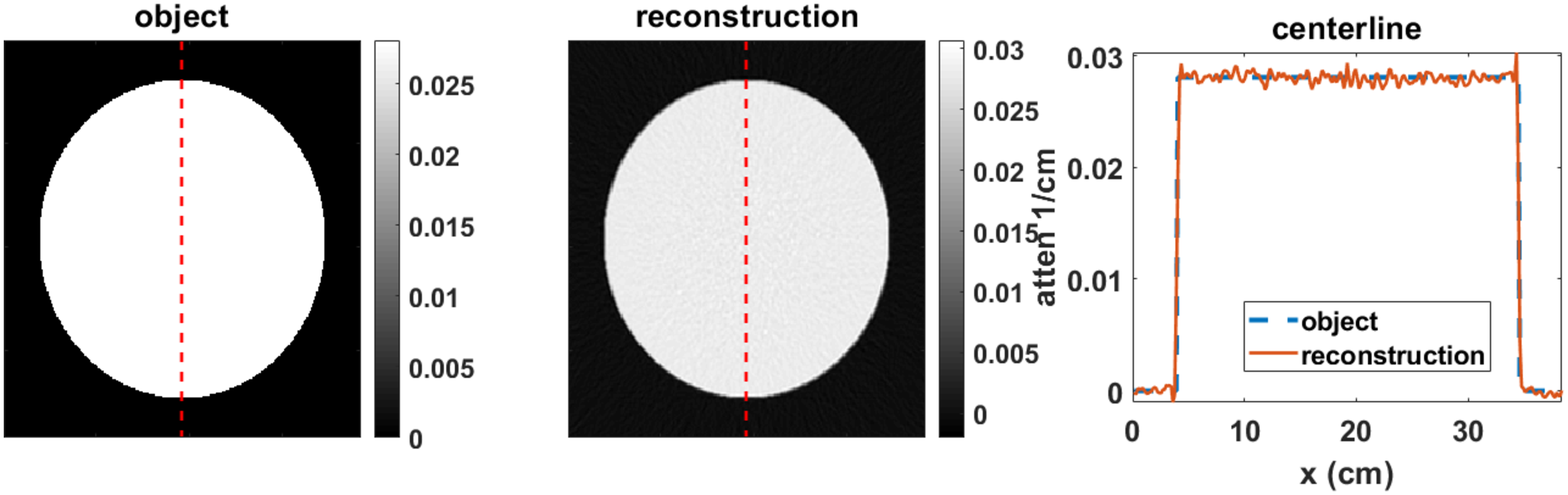}
\caption{Reconstruction of the circular phantom. The object $\mu(E,\bm R)$ (left), the reconstruction $\hat{\mu}(E,\bm R)$ (middle) and the center-line plots (right) are all presented at $E=75.5$~keV. }
\label{fig:water_phantom_recon}
\end{figure}

\indent The first phantom reconstructed is a circular water phantom of diameter about 30~cm with pixel-pitch 0.15~cm. This phantom and its reconstruction are plotted at $E=75.5$~keV in Figure~\ref{fig:water_phantom_recon}. The reconstruction matches well with the object. As shown in the center-line plot in the right panel of Figure~\ref{fig:water_phantom_recon}, the reconstruction has no cupping artifacts, which is typically associated with beam hardening. 

\begin{figure}[htbp!]
\centering
\includegraphics[width=\linewidth]{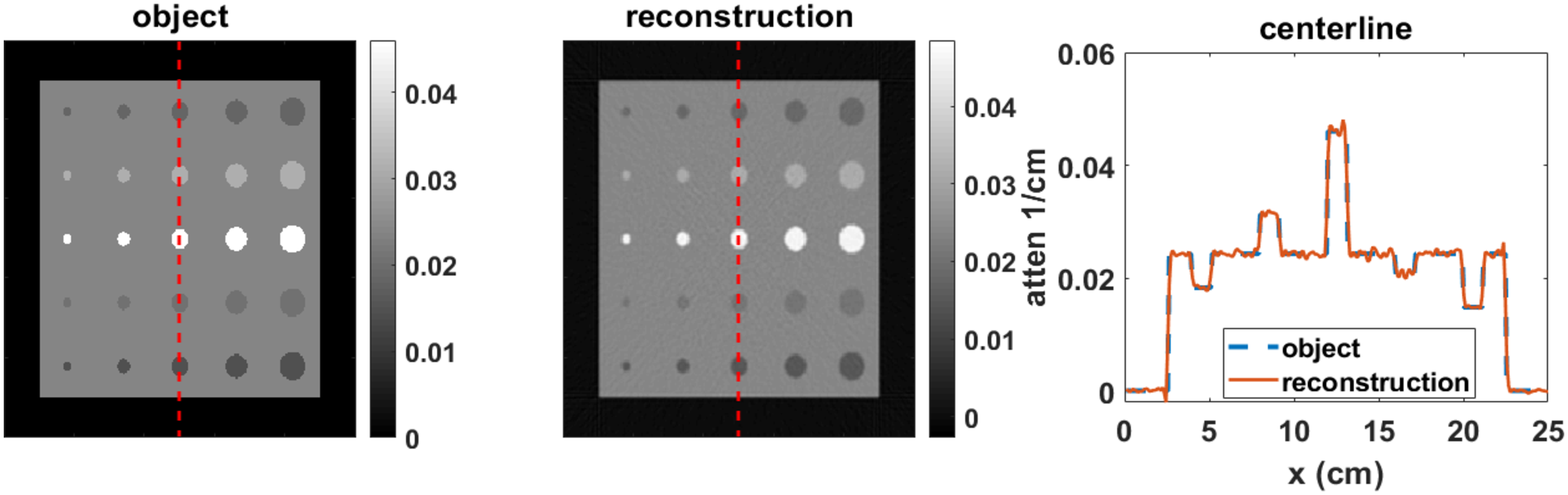}
\caption{Reconstruction of the resolution phantom. The object $\mu(E,\bm R)$ (left), the reconstruction $\hat{\mu}(E,\bm R)$ (middle) and the center-line plots (right) are all presented at $E=75.5$~keV. }
\label{fig:duke_phantom_recon}
\end{figure}

\indent  A second phantom is a multi-material resolution phantom design inspired by Gong et.al. \cite{gong2018rapid}. The length of the phantom is 20~cm with pixel pitch around 0.1~cm. The phantom, as shown in Figure~\ref{fig:duke_phantom_recon} (left), is a Delrin block with 25 circular inserts in five rows and five columns. In each row, the inserts are made from the same material; in each column, the inserts have the same diameter. From top to bottom, the five materials for the inserts are water, polyvinyl chloride, cast magnesium, acrylic and methanol. The diameter of the inserts are from 0.6 to 1.8 cm with 0.3 cm step. The reconstruction of the resolution phantom are plotted at $E=75.5$~keV in Figure~\ref{fig:duke_phantom_recon}. In the plots, different shades of grey represent different materials. The reconstruction results show that the ML estimation algorithm works for a broad range of AM coefficients 

\indent Each reconstruction, which calls the ML estimation algorithm 88,200 times, takes approximately 130 seconds on a desktop with a quad-core central processing unit (CPU). The reconstruction can be further sped up using a graphic processing unit (GPU).

%%%%%%%%%%%%%%%%%% discussion %%%%%%%%%%%%%%%%%%%%
\section{Discussion} \label{sec:disc}

\indent In our proof of invertibility, we focused on the interior points of $M_1$ and $M_2$ and proved that, for an ME X-ray transform, the global invertibility is equivalent to local invertibility for $\bm A$ in the interior of $M_1$ and $\bm I$ in the interior of $M_2$. This equivalence can be extended to the boundaries of $M_1$ and $M_2$ by invoking Theorem 6 in Sandberg et. al. \cite{sandberg1980global}. As mentioned in the introduction, Bal et. al. \cite{bal2020uniqueness} has also provided a sufficient condition for the invertibility of ME X-ray transform. Their sufficient condition is that the Jacobian is a P-matrix in a rectangle in $\bm{R}^N$. P-matrix, which is a concept related to the preservation of orientation, requires the matrix and a few sub-matrices to be all positive. For the definition of P-matrix, please refer to Bal et. al. \cite{bal2020uniqueness} or Gale and Nikaido \cite{gale1965jacobian}. Bal et. al. studied the invertibility of different ME X-ray systems by numerically calculating the Jacobian matrix on a grid of $\bm A$ values for each system. Based on numerical simulations, they have suggested that an ME X-ray system may become invertible as soon as the mapping is locally invertible in the rectangle. Our sufficient condition for global invertibility is that the Jacobian is non-vanishing for all points in $M_1$. In comparison to Bal et. al., our sufficient condition is weaker (better), but the domain where the Jacobian matrix needs to be checked is larger. For ease of computation, we also provide a sufficient condition for non-vanishing Jacobian, which requires the integrand of the Jacobian to have constant sign over all energy combinations in $S_{12...N}$. From a practical point of view, the latter condition is significantly easier to use, as (1) the sign of the Jacobian integrand does not depend on $\bm A$, and (2) the properties of the basis functions  and the detector response functions can be studied separately. 

\indent Invertibility only requires that the Jacobian $J(\bm A)\neq0$ for all coefficients $\bm A$. Nonetheless, a smaller $J(\bm A)$ leads to a worse-conditioned inverse problem and hence more uncertainty in the estimation as discussed in Section~\ref{sec:crb}. Take the binning detector depicted in Figure~\ref{fig:det_illu} (c) as an example, when Bin 1 and Bin 3 do not overlap, the system is invertible for $N=3$ (when imaging materials with no K-edges). However, the overlap between bins would result in a reduction in the Jacobian and hence more uncertainty in the coefficient estimation, which has also been observed in other works \cite{ren2021conditioning, tangconditioning}. One can employ the CRLB to optimize bin boundaries for a given set of AM coefficients $\bm A$. As the CRLB varies with $\bm A$, the optimum bin boundaries depend on the prior distribution of the objects. An optimum energy-weighting strategy that is not object dependent has been proposed by Wang et.\ al.\ \cite{wang2010sufficient}. Their strategy is to set the weights $p_m(E)$ same as the attenuation basis functions $f_i(E)$. They have proved that this measurement strategy provided a sufficient statistic to the X-ray spectral flux. From Equation~(\ref{eq:jacobian_subspace}), we can prove that this strategy is globally invertible, because $\det[F(\bm E)]\det[P(\bm E)]=\{\det[F(\bm E)]\}^2\geq0$, for all $\bm E$. 

\indent Conventional DECT systems reconstruct the effective atomic number ($Z_{e}$) and the electron density ($\rho$) \cite{martz2016ct} from two energy-weighted measurements. However, $Z_{e}$ and $\rho$ may not capture all of the information about material composition measurable from attenuation-based X-ray systems. Based on principal component analysis (PCA), Bornefalk et.\ al.\ \cite{bornefalk2012xcom} have suggested that the intrinsic dimensionality of the attenuation profiles of low-Z materials in the XCOM data base is four. Midgley et.\ al.\ \cite{midgley2004parameterization}  also showed similar degrees of freedom in the parameterization of the X-ray linear attenuation profiles. However, whether these intrinsic dimensions are accessible or not is still up to debate \cite{midgley2005materials, levine2018preliminary}. There is potential value in collecting ME X-ray data, but the benefits may depend on the task of the imaging system and the experimental setup. 

\indent We used a set of basis functions that describe photoelectric, Compton scattering and Rayleigh scattering, because we wanted to investigate whether the Rayleigh coefficient is estimable or not. Rayleigh scattering has often been ignored in DE imaging due to its small contribution in the X-ray attenuation profile. Our results show that the Rayleigh component, $A_{rs}$, is solvable and the uncertainty in its estimation is comparable to that of the photoelectric coefficient. However, we did not specifically study how important $A_{rs}$ is for the task of material discrimination. Other basis functions that are based on materials of interest (such as water and bone) or on PCA \cite{alvarez2013dimensionality, bornefalk2012xcom, weaver1985attenuation, xie2020principal} can be used as well. As pointed out by Alvarez et.\ al.\ \cite{alvarez2019invertibility}, the choice of a particular basis set does not affect the invertibility. The uncertainty in the estimated attenuation profile should not be affected by the choice of the basis set either.  

\indent We demonstrated a two-step reconstruction algorithm that consists of an ML estimation of the AM coefficients and the FBP reconstruction. Many work have been done in DECT and MECT reconstruction. Reconstruction algorithms are currently available in three main flavors: object-domain based \cite{brooks1977quantitative, maass2009image}, projection-domain based \cite{abascal2018nonlinear, wu2016weighted}, and one-step statistical algorithms \cite{mory2018comparison, mechlem2017joint, long2014multi, barber2016algorithm, kazantsev2018joint} that estimate $\hat{\bm a}(\bm R)$ from the raw data directly. Our ML estimation algorithm was designed for Poisson likelihood and ideal ME X-ray transform where Equation~(\ref{eq:mean_mapping}) is valid. When the Poisson likelihood model or the ideal forward model are not accurate, the estimation algorithm needs modification. 

\indent Our ML estimation algorithm is almost unbiased and achieves the Cr\'amer-Rao lower bound. The ML estimation is an established paradigm for nonlinear estimation tasks. At high signal-to-noise ratio (SNR), ML estimates are asymptotically unbiased and efficient (achieves Cr\'amer-Rao lower bound). At low SNR (e.g. short exposure time), however, the ML estimates tend to be skewed and the variance is often larger than the Cramer-Rao bound \cite{mueller1995estimation}. The reason why our estimator is efficient is probably that our simulation was carried out in the high SNR regime. In our experiment, the smallest photon count collected in an energy bin is about 300 photons, which still has a relatively high SNR. To analyze a realistic system, one need to first identify if the system is operating at low SNR regime. If that is the case, instead of using the Cr\'amer-Rao bound to characterize the variance of the ML estimates, one can apply other measures such as $\chi^2_{pdf-ML}$-isocontours \cite{muller2005measures} to describe the distribution of the ML estimates. 

\indent There are several limitations in our work. Firstly, the physical process considered by the mapping $\mathcal{X}: {\bm A}\rightarrow {\bm I}$ includes only the attenuation of the X-ray photons, which follows Beer's law, but not signals due to scattered radiation or background radiation. Although scattered radiation and background radiation can be significantly mitigated through anti-scatter grids \cite{tang1998anti}, those signals should be characterized and accounted for, as they may affect the invertibility of a realistic system. Secondly, effects that limit detector performance, such as charge sharing \cite{shikhaliev2009photon, xu2011evaluation}, charge trapping \cite{knoll2010radiation,xu2011evaluation} and pulse pileup, are ignored. Such details should be considered in the system model when the framework is applied to a specific imaging system. For a given realistic detector response function, the invertibility of the system can be studied by calculating $\det[P({\bm E})]$ numerically over the subspace $\Omega_{12…N}$. In this case, $\det[P({\bm E})]\leq0$ (or $\geq0$) over $\Omega_{12…N}$ may not be guaranteed, hence the invertibility is not guaranteed. However, one can apply the invertibility framework in the optimization of detector designs. Lastly, we assumed the energy-weighting functions, including the source spectrum and the detector energy response functions, are known exactly. In reality, one can measure the energy-weighting functions experimentally \cite{ha2019estimating} within some uncertainty. 

%\indent Another limitation is that, in the derivation of the CRB bound, we assumed the measurement data follow Poisson distribution. For many realistic detectors, the Poisson model is not an accurate model for the measurement data. For example, the energy-weighting functions of a CIX-PC/EI/MI detector introduce correlations between different measurements; and the data collected by a PC detector also have correlations partially due to pulse pileup. Furthermore, electronic noise and nonlinearities may also affect the statistics of the data. Finding the optimum statistical model to describe measurement data for a specific detector is an interesting avenue for future work.

%%%%%%%%%%%%%%%%%% conclusion %%%%%%%%%%%%%%%%%%%%

\section{Conclusion}

\indent We have provided a sufficient condition for the global invertibility of an ME X-ray transform for attenuation-based X-ray imaging. The ME X-ray transform is defined as the mapping from $N$ ($N\geq2$) AM coefficients to $N$ energy-weighted noise-free measurements. The invertibility of this transform depends greatly on the weighting schemes used in the measurements. Considering scenes with no K-edge materials, we represented the X-ray attenuation profiles with $N=3$ AM coefficients and proved the global invertibility of the transform for four commonly used weighting schemes. The same framework can be used to examine the invertibility of an arbitrary ME X-ray system, such as a system with non-ideal detectors, a system with multiple source emission spectra, and scenes with K-edge materials. This mathematical framework can be applied broadly in the design of X-ray detectors and systems. 

\indent We also considered Poisson noise in the measurement data and presented the CRLB on the estimation uncertainty. Furthermore, we presented an ML estimation algorithm and applied the algorithm to estimate AM coefficients for varying lengths of water and varying lengths of iron. The results have shown that the coefficient corresponding to Rayleigh scattering is estimable. Last but not least, we demonstrated the application of the ML estimator in reconstruction. Two phantoms imaged through a simulated fan-beam CT with ideal three-energy discriminating photon-counting detectors were reconstructed. The reconstructed images match well with the objects and are free of the `cupping artifacts' induced by beam hardening.

%%%%%%%%%%%%%%%%%% declarations %%%%%%%%%%%%%%%%%%%%

\section*{Acknowledgments}
Dr. Clarkson acknowledges the support of NIH R01- EB000803 and P41- EB002035.
%The authors gratefully acknowledge the support of the U.S. Department of Homeland Security (DHS). The research for this project was conducted under contract with the DHS Science and Technology Directorate (S\&T), contract HSHQDC-16-C-B0014. The opinions contained herein are those of the contractors and do not necessarily reflect those of DHS S\&T.

\section*{Disclosures}
The authors declare that there are no conflicts of interest related to this article.

\section*{Data Availability Statement}
The data and code that support the findings of this study are available from the corresponding author upon reasonable request.

%%%%%%%%%%%%%%%% References %%%%%%%%%%%%%%%%
\section*{References}
%%%% AMA style
%\bibliographystyle{ama}
%\bibliography{Ebins}

%%%%%%%%%%%%% begin .bib file %%%%%%%%%%%%%%%%%%%%%%

%%%%%%%%%%%%% end .bib file %%%%%%%%%%%%%%%%%%%%%%

\section*{Figures}

\end{document}